\documentclass[aps,prd,twocolumn,showpacs,floats,floatfix,letterpaper,nofootinbib,superscriptaddress,]{revtex4}

\usepackage{amssymb,amsmath,latexsym,mathrsfs}
\usepackage{graphicx}
\usepackage{epsfig}
\usepackage{multirow}
\usepackage{array}
\usepackage{color}

\newcommand{\bea}{\begin{eqnarray}}
\newcommand{\eea}{\end{eqnarray}}

\begin{document}


\title{Cosmological Hints of Modified Gravity ?}

\author{Eleonora Di Valentino}
\affiliation{Institut d'Astrophysique de Paris (UMR7095: CNRS \& UPMC- Sorbonne Universities), F-75014, Paris, France}
\author{Alessandro Melchiorri}
\affiliation{Physics Department and INFN, Universit\`a di Roma ``La Sapienza'', Ple Aldo Moro 2, 00185, Rome, Italy}
\author{Joseph Silk}
\affiliation{Institut d'Astrophysique de Paris (UMR7095: CNRS \& UPMC- Sorbonne Universities), F-75014, Paris, France}
\affiliation{AIM-Paris-Saclay, CEA/DSM/IRFU, CNRS, Univ. Paris VII, F-91191 Gif-sur-Yvette, France}
\affiliation{Department of Physics and Astronomy, The Johns Hopkins University Homewood Campus, Baltimore, MD 21218, USA}
\affiliation{BIPAC, Department of Physics, University of Oxford, Keble Road, Oxford
OX1 3RH, UK}

\begin{abstract}
The recent measurements of Cosmic Microwave Background temperature and polarization anisotropies made by the 
Planck satellite have provided impressive confirmation of the $\Lambda$CDM cosmological model.
However  interesting hints of slight deviations from $\Lambda$CDM have been found, including a $95 \%$ c.l. preference for a "modified gravity" structure formation scenario. In this paper we confirm the preference for a modified gravity scenario from Planck 2015 data,
find that modified gravity solves the so-called $A_{lens}$ anomaly in the CMB angular spectrum, and constrains the  amplitude of matter density fluctuations  to $\sigma_8=0.815_{-0.048}^{+0.032}$,
in better agreement with weak lensing constraints.
Moreover, we find a lower value for the reionization optical depth of $\tau=0.059\pm0.020$
(to be compared with the value of $\tau= 0.079 \pm 0.017$  obtained in the standard scenario), more consistent 
with recent optical and UV data.  We check the stability of this result by considering possible degeneracies with other parameters, including  the neutrino effective number, the running of the spectral index and the amount of primordial helium. 
The indication for modified gravity is still present at about $95\%$ c.l., and could become more significant if 
lower values of $\tau$ were to be further confirmed by future cosmological and astrophysical data.
When the CMB lensing likelihood is included in the analysis the statistical significance for MG simply vanishes,
indicating also the possibility of a systematic effect for this MG signal. 
\end{abstract}


\maketitle

\section{Introduction}

The recent measurements of Cosmic Microwave Background (CMB) anisotropies by the Planck satellite experiment
\cite{planckmain,planckparams2015} have fully confirmed, once again, the expectations of the standard cosmological model based on cold dark matter, inflation and a cosmological constant. 

While the agreement is certainly impressive, some hints for deviations from the standard scenario
have emerged that certainly deserve further investigation.
In particular, an interesting hint for "modified gravity" (MG hereafter), i.e. a deviation of the growth
of density perturbations from that expected under General Relativity (GR hereafter),
 has been reported in \cite{Ade:2015rim} using a phenomenological parametrization 
to characterize non-standard metric perturbations.

In  past years, several authors (see e.g. \cite{Zhao:2010dz,Daniel:2010ky,Simpson:2012ra,DiValentino:2012yg,Giannantonio:2009gi,Macaulay:2013swa,
Hojjati:2013xqa,Marchini:2013oya,Hu:2013aqa,Munshi:2014tua,Boubekeur:2014uaa,Hu:2014sea, Cataneo:2014kaa,Dossett:2015nda,Johnson:2015aaa,Ade:2015rim}) have constrained possible deviations of the evolution of perturbations with respect to the $\Lambda$CDM model, by parametrizing  the gravitational potentials $\Phi$ and $\Psi$ and their linear combinations.
Considering the parameter $\Sigma$, that modifies the
lensing/Weyl potential given by the sum of the Newtonian and curvature 
potentials $\Psi+\Phi$, the analysis of \cite{Ade:2015rim} reported the
current value of $\Sigma_0-1=0.28\pm0.15$ at $68 \%$ from Planck CMB temperature data, i.e. a 
deviation from the expected GR null value at about two standard deviations.
The discrepancy with GR increases when weak lensing data is included, bringing the
constrained value to $\Sigma_0-1=0.34^{+0.17}_{-0.14}$ (again, see \cite{Ade:2015rim}).

This result is clearly interesting and should be further investigated. Small systematics may still
certainly be present in the data and a further analysis, expected by $2016$, 
from the Planck collaboration could solve the issue.
In the meantime, it is certainly timely to independently reproduce the result presented in \cite{Ade:2015rim}
and to investigate its robustness, especially in view of other anomalies and tensions currently
present in cosmological data.

Indeed, another anomaly seems to be suggested by the Planck data, i.e. the amplitude of
gravitational lensing in the angular spectra. This quantity, parametrized 
by the lensing amplitude $A_{\rm lens}$ as firstly introduced in \cite{erminia},
is also larger than expected at the level of two standard deviations.
The Planck+LowP analysis of \cite{planckparams2015} reports the value of
$A_{\rm lens}=1.22 \pm 0.10$ at $68 \%$ c.l.. 
This anomaly persists even when considering a significantly extended parameter space
as shown in \cite{DiValentino:2015ola}.
It is therefore mandatory to check if this deviation is in some way connected with the "$\Sigma_0$" anomaly
performing an analysis by varying both parameters at the same time.
This has been suggested but not actually done in \cite{Ade:2015rim}.

Moreover, some  mild tension seems also to be present between the large angular scale
Planck LFI polarization data (that, alone, provides a constraint on the
optical depth $\tau = 0.067 \pm 0.023$ \cite{planckparams2015}) and the Planck HFI small-scale temperature
and polarization data that, when combined with large-scale LFI polarization,
shifts the constraint to  $\tau = 0.079 \pm 0.017$ \cite{planckparams2015}.
Since the Planck constraints on $\tau$ are model-dependent, is meaningful to check
if the assumption of MG could, at least  partially, resolve the "$\tau$" tension.

Another tension concerns the amplitude of the r.m.s. density fluctuations on scales of $8$ Mpc $h^{-1}$, the so-called $\sigma_8$ parameter. The constraints on $\sigma_8$ derived by the Planck data under the assumption of GR and $\Lambda$-CDM are in tension with the same quantity observed by low redshift surveys based on clusters counts, lensing and
redshift-space distortions (see e.g. \cite{battye} and \cite{planckparams2015}).
This tension appears most dramatic when considering the weak lensing measurements provided by
the CFHTLenS survey (see discussion in \cite{Ade:2015rim}), which prefer lower values of $\sigma_8$ 
with respect to those obtained by Planck.  Several solutions to this mild tension have been proposed, including  dynamical dark energy
\cite{kunz}, decaying dark matter \cite{enqvist,dolgov}, ultralight axions \cite{hlozek}, and voids \cite{ichiki}.
It is therefore  timely to further check if the "$\sigma_8$ tension" could be
reconciled by assuming MG. This approach has already been suggested, for example,
by \cite{Dossett:2015nda}.

Finally, there are also extra parameters such as  the running of the spectral index $dn_S/dlnk$,
the neutrino effective number $N_{eff}$ (see e.g. \cite{neff}),  and the
helium abundance $Y_p$ (see e.g. \cite{Trotta:2003xg}) that could be varied and that could in principle
 be correlated with MG. Since the values of these parameters derived under $\Lambda$-CDM (see \cite{planckparams2015})
 are consistent with standard expectations, it is crucial to investigate  whether the inclusion of MG could change these conclusions.

This paper is organized as follows: in the next section we describe the  MG parametrization 
that we consider, while in Section III we describe the data analysis method adopted. In Section IV, we present our results
and in Section V we derive our conclusions.

\section{Perturbation equations}

Let us briefly explain here how MG is implemented in our analysis, discussing the relevant equations. 
Assuming a flat universe, we can write the line element of the Friedmann-Lemaitre-Robertson-Walker (FLRW) metric 
in the conformal Newtonian gauge as:
\begin{equation}  
\label{metric}
ds^2 = a(\tau)^2[-(1+2\Psi)d\tau^2+(1-2\Phi)dx^idx_i] \,,
\end{equation}
where $a$ is the scale factor, $\tau$ is the conformal time, $\Psi$ is the Newton's gravitational potential, and $\Phi$ the space curvature \footnote{In the synchronous gauge, that is the one adopted in boltzmann codes as  \texttt{CAMB} \cite{Lewis:1999bs}, we have:
\begin{equation}  
ds^2 = a(\tau)^2[-d\tau^2+(\delta_{ij} + h_{ij})dx^i dx^j] \,,
\end{equation}
where $h_{ij}$ are defined as in \cite{Ma:1995ey}.}.

Given the line element of the Eq.~\ref{metric}, we can use a phenomenological parametrization of the gravitational potentials $\Psi$ and $\Phi$ and their combinations. We consider the parametrization used in the publicly available code \texttt{MGCAMB} \cite{Zhao:2008bn,Hojjati:2011ix}, introducing the scale-dependent function $\mu(k,a)$, that modifies the Poisson equation for $\Psi$:
\begin{equation}  
k^2\Psi=-4 \pi G a^2\mu(k,a)\rho\Delta \,,
\end{equation}
where $\rho$ is the dark matter energy density, $\Delta$ is the comoving density perturbation. Furthermore one can consider the function $\eta(k,a)$, that takes into account the presence of a non-zero anisotropic stress:
\begin{equation}  
\eta(k,a)=\frac{\Phi}{\Psi} \,.
\end{equation}
We can then easily introduce the function $\Sigma(k ,a)$, which modifies the lensing/Weyl
potential $\Phi+\Psi$ in the following way:
\begin{equation}
 -k^2 (\Phi+\Psi) \equiv 8\pi G a^2 \Sigma(k,a) \rho  \Delta \,,
\end{equation}
and that can be obtained directly from $\mu(k,a)$ and $\eta(k,a)$ as 
\begin{equation} 
 \Sigma=\frac{\mu}{2} (1+\eta ) \,.
\end{equation}
Of course, if we have GR then $\mu = \eta = \Sigma=1$.

It is now useful to give an expression for $\mu$ and $\eta$. Following Ref.\cite{Ade:2015rim}, we parametrize $\mu$ and $\eta$ as:
\begin{eqnarray}
\mu(k,a) &=& 1 + f_1(a) \frac{1 + c_1 (\lambda H/k)^2}{1+(\lambda H/k)^2}; \\
\eta(k,a) &=& 1 + f_2(a) \frac{1 + c_2 (\lambda H/k)^2}{1+(\lambda H/k)^2},
\end{eqnarray}
where $H=\dot a /a$ is the Hubble parameter, $c_1$ and $c_2$ are constants and the 
$f_{\rm{i}}(a)$ are functions of time that characterize the amplitude of the deviation from GR.

Again, following \cite{Ade:2015rim} we choose a time dependence for these functions
related to the dark energy density:

\begin{equation}
f_{\rm{i}}(a) = E_{\rm{ii}} \Omega_{\rm{DE}}(a)\,,
\end{equation}

where  $E_{\rm{ii}}$ are, again, constants and $\Omega_{\rm{DE}}(a)$ is the dark energy density parameter.
As discussed in  Ref.~\cite{Ade:2015rim}, the inclusion of scale dependence does not change significantly the
results, we can therefore consider the scale independent parametrization, in which $c_1=c_2=1$. 
In other words, we modify the publicly available code \texttt{MGCAMB} \cite{Zhao:2008bn,Hojjati:2011ix}, by substituting to the original $\mu$ and $\eta$, the following parametrizations:

\begin{eqnarray} \label{DErel}
\mu(k,a) &=& 1 + E_{\rm{11}} \Omega_{\rm{DE}}(a)\,; \\
\eta(k,a) &=& 1 + E_{\rm{22}} \Omega_{\rm{DE}}(a)\,.
\end{eqnarray}

A detection of $E_{\rm{ii}} \neq 0$ could therefore indicate a departure of the evolution of density
perturbations from GR. In order to further simplify the problem, we assume a cosmological
constant for the background evolution.

\section{Method}

\begin{table}
\begin{center}
\begin{tabular}{c|c}
Parameter                    & Prior\\
\hline
$\Omega_{\rm b} h^2$         & $[0.005,0.1]$\\
$\Omega_{\rm c} h^2$       & $[0.001,0.99]$\\
$\Theta_{\rm s}$             & $[0.5,10]$\\
$\tau$                       & $[0.01,0.8]$\\
$n_s$                        & $[0.8, 1.2]$\\
$\log[10^{10}A_{s}]$         & $[2,4]$\\
$E_{11}$         & $[-1,3]$\\
$E_{22}$         & $[-1.4,5]$\\
$\frac{dn_s}{dlnk}$ & [-1,1]\\
$N_{\rm eff}$ & [0.05,10]\\
$A_{\rm lens}$ & [0,10]\\
$Y_P$& [0.1,0.5]\\
\end{tabular}
\end{center}
\caption{External flat priors on the cosmological parameters assumed in this paper.}
\label{priors}
\end{table}

We consider flat priors listed in Table~\ref{priors} on all the parameters that we are constraining. They are: the six parameters of the $\Lambda$CDM model, i.e. 
the Hubble constant $H_0$, the baryon $\Omega_{\rm b} h^2$ and 
cold dark matter  $\Omega_{\rm c} h^2$ energy densities, the primordial amplitude and spectral index 
of scalar perturbations, $A_s$ and $n_s$ respectively, (at pivot scale $k_0=0.05 hMpc^{-1}$), 
and the reionization optical depth $\tau$; the constant parameters of MG, $E_{11}$ and $E_{22}$;
the several extensions to $\Lambda$CDM model. In particular we vary the neutrino effective number $N_{\textrm{eff}}$ (see e.g. \cite{neff}), the running of the scalar spectral index $dn_S /dlnk$, the primordial Helium abundance $Y_P$
and the lensing amplitude in the angular power spectra $A_{\rm lens}$.
We also vary foreground parameters following the same method of  \cite{plancklike2015} and \cite{planckparams2015}.

\begin{table*}[t]
\begin{center}
\scalebox{1}{\begin{tabular}{ccccccc}
\hline\hline
\hspace{1mm}\\
& Planck TT & Planck TT + WL & Planck TT + lensing & Planck pol & Planck pol + WL & Planck pol + lensing \\
\hline\hline
\hspace{1mm}\\
$E_{11}$ & $0.08^{+0.33}_{-0.72}$ & $-0.18^{+0.19}_{-0.49}$ & $0.08^{+0.34}_{-0.59}$ & $0.06^{+0.33}_{-0.66}$ & $-0.21^{+0.19}_{-0.45}$ & $0.08^{+0.35}_{-0.54}$ \\
\hspace{1mm}\\
$E_{22}$& $1.0^{+1.3}_{-1.6}$ & $1.9^{+1.4}_{-1.0}$ & $0.4^{+0.9}_{-1.4}$ & $0.9^{+1.2}_{-1.5}$ & $1.7^{+1.3}_{-1.0}$ & $0.4^{+0.8}_{-1.3}$ \\
\hspace{1mm}\\
$\mu_0 -1$& $0.05^{+0.23}_{-0.50}$ & $-0.13^{+0.13}_{-0.35}$ & $0.05^{+0.24}_{-0.41}$ & $0.04^{+0.23}_{-0.45}$ & $-0.15^{+0.13}_{-0.32}$ & $0.05^{+0.24}_{-0.38}$\\
\hspace{1mm}\\
$\eta_0 -1$& $0.7^{+0.9}_{-1.2}$ & $1.3^{+1.0}_{-0.7}$ & $0.31^{+0.61}_{-0.94}$ & $0.6^{+0.8}_{-1.0}$ & $1.20^{+0.91}_{-0.68}$ & $0.26^{+0.56}_{-0.86}$\\
\hspace{1mm}\\
$\Sigma_0 -1$& $0.28 \pm 0.15$ & $0.34^{+0.16}_{-0.15}$ & $0.11^{+0.09}_{-0.12}$ & $0.23 \pm 0.13$ & $0.27 \pm 0.13$ & $0.10^{+0.09}_{-0.11}$\\
\hspace{1mm}\\
\hline
\hspace{1mm}\\
$\Omega_{\rm b}h^2$& $0.02251 \pm 0.00027$ & $0.02263 \pm 0.00026$ & $0.02238 \pm 0.00024$ & $0.02237 \pm 0.00017$ & $0.02243 \pm 0.00017$ & $0.02233 \pm 0.00016$ \\
\hspace{1mm}\\
$\Omega_{\rm c}h^2$ & $0.1175 \pm 0.0024$ & $0.1159 \pm 0.0022$ & $0.1171 \pm 0.0021$ & $0.1188 \pm 0.0016$ & $0.1180 \pm 0.0015$ & $0.1185 \pm 0.0014$ \\
\hspace{1mm}\\
$H_0$ & $68.5 \pm 1.1$ & $69.2 \pm 1.1$ & $68.47 \pm 0.99$ & $67.78 \pm 0.71$ & $68.15 \pm 0.69$ & $67.83 \pm 0.66$\\
\hspace{1mm}\\
$\tau$ & $0.065 \pm 0.021$ & $0.061^{+0.020}_{-0.023}$ & $0.050 \pm 0.019$ & $0.059 \pm 0.020$ & $0.054 \pm 0.019$ & $0.045 \pm 0.017$\\ 
\hspace{1mm}\\
$n_s$ & $0.9712 \pm 0.0071$ & $0.9754 \pm 0.0067$ & $0.9706 \pm 0.0062$ & $0.9668 \pm 0.0051$ & $0.9689 \pm 0.0050$ & $0.9668 \pm 0.0047$\\
\hspace{1mm}\\
$\sigma_8$ & $0.816 ^{+0.034}_{-0.052}$ & $0.787 ^{+0.022}_{-0.039}$ & $0.802 ^{+0.033}_{-0.039}$ & $0.815 ^{+0.032}_{-0.048}$ & $0.788 ^{+0.021}_{-0.035}$ & $0.803 \pm 0.031$\\
\hspace{1mm}\\
\hline\hline
\end{tabular}}
\caption{Constraints at $68 \%$ c.l. on the cosmological parameters assuming modified gravity (parametrized by
$E_{11}$ and $E_{22}$) and varying the $6$ parameters
of the standard $\Lambda$CDM model.}
\label{table2}
\end{center}
\end{table*}

\begin{table*}[t]
\begin{center}
\scalebox{1}{\begin{tabular}{ccccccc}
\hline\hline
\hspace{1mm}\\
& Planck TT & Planck TT + WL & Planck TT + lensing & Planck pol & Planck pol + WL & Planck pol + lensing \\
\hline\hline
\hspace{1mm}\\
$E_{11}$ & $0.06^{+0.33}_{-0.65}$ & $-0.15^{+0.22}_{-0.51}$ & $0.08^{+0.33}_{-0.63}$ & $0.07^{+0.33}_{-0.62}$ & $-0.18^{+0.21}_{-0.47}$ & $0.06^{+0.33}_{-0.63}$ \\
\hspace{1mm}\\
$E_{22}$& $0.8^{+1.1}_{-1.7}$ & $1.4^{+1.4}_{-1.3}$ & $0.8^{+1.0}_{-1.5}$ & $0.7^{+1.0}_{-1.6}$ & $1.4 \pm 1.2$ & $0.8^{+1.1}_{-1.6}$ \\
\hspace{1mm}\\
$\mu_0 -1$& $0.04^{+0.23}_{-0.46}$ & $-0.10^{+0.15}_{-0.36}$ & $0.06^{+0.23}_{-0.44}$ & $0.05^{+0.23}_{-0.43}$ & $-0.12^{+0.15}_{-0.33}$ & $0.04^{+0.22}_{-0.44}$\\
\hspace{1mm}\\
$\eta_0 -1$& $0.6^{+0.7}_{-1.2}$ & $1.0^{+1.0}_{-0.9}$ & $0.5^{+0.7}_{-1.1}$ & $0.5^{+0.7}_{-1.1}$ & $0.95 \pm 0.81$ & $0.6^{+0.7}_{-1.1}$\\
\hspace{1mm}\\
$\Sigma_0 -1$& $0.21 ^{+0.16}_{-0.21}$ & $0.22^{+0.17}_{-0.22}$ & $0.21^{+0.15}_{-0.17}$ & $0.19^{+0.14}_{-0.18}$ & $0.20^{+0.14}_{-0.18}$ & $0.22^{+0.14}_{-0.16}$\\
\hspace{1mm}\\
\hline
\hspace{1mm}\\
$\Omega_{\rm b}h^2$& $0.02259 \pm 0.00029$ & $0.02273 \pm 0.00028$ & $0.02231 \pm 0.00026$ & $0.02239 \pm 0.00017$ & $0.02246 \pm 0.00017$ & $0.02229 \pm 0.00016$ \\
\hspace{1mm}\\
$\Omega_{\rm c}h^2$ & $0.1169 \pm 0.0025$ & $0.1152 \pm 0.0023$ & $0.1180 \pm 0.0025$ & $0.1187 \pm 0.0016$ & $0.1177 \pm 0.0015$ & $0.1191 \pm 0.0015$ \\
\hspace{1mm}\\
$H_0$ & $68.8 \pm 1.2$ & $69.6 \pm 1.1$ & $68.1 \pm 1.2$ & $67.82 \pm 0.73$ & $68.26 \pm 0.69$ & $67.59 \pm 0.70$\\
\hspace{1mm}\\
$\tau$ & $0.059^{+0.021}_{-0.023}$ & $0.054 \pm 0.021$ & $0.059 \pm 0.021$ & $0.056 \pm 0.020$ & $0.049^{+0.019}_{-0.022}$ & $0.057 \pm 0.021$\\ 
\hspace{1mm}\\
$n_s$ & $0.9730 \pm 0.0073$ & $0.9772 \pm 0.0068$ & $0.9687 \pm 0.0070$ & $0.9671 \pm 0.0050$ & $0.9694 \pm 0.0049$ & $0.9656 \pm 0.0049$\\
\hspace{1mm}\\
$\sigma_8$ & $0.807 ^{+0.033}_{-0.049}$ & $0.782 ^{+0.025}_{-0.038}$ & $0.813 ^{+0.033}_{-0.046}$ & $0.813 ^{+0.032}_{-0.044}$ & $0.786 ^{+0.023}_{-0.035}$ & $0.814^{+0.031}_{-0.046}$\\
\hspace{1mm}\\
$A_{\rm lens}$& $1.09^{+0.10}_{-0.13}$ & $1.13^{+0.10}_{-0.14}$ & $0.924^{+0.065}_{-0.089}$ & $1.04^{+0.08}_{-0.10}$ & $1.07^{+0.09}_{-0.11}$ & $0.914^{+0.062}_{-0.078}$ \\
\hspace{1mm}\\
\hline\hline
\end{tabular}}
\caption{Constraints at $68 \%$ c.l. on the cosmological parameters assuming modified gravity (parametrized by
$E_{11}$ and $E_{22}$) and varying the $6$ parameters
of the standard $\Lambda$CDM model plus $A_{\rm lens}$.}
\label{table3}
\end{center}
\end{table*}

\begin{table*}[t]
\begin{center}
\scalebox{1}{\begin{tabular}{ccccccc}
\hline\hline
\hspace{1mm}\\
& Planck TT & Planck TT + WL & Planck TT + lensing & Planck pol & Planck pol + WL & Planck pol + lensing \\
\hline\hline
\hspace{1mm}\\
$E_{11}$ & $0.07^{+0.31}_{-0.73}$ & $-0.13^{+0.20}_{-0.58}$ & $0.09^{+0.35}_{-0.64}$ & $0.07^{+0.34}_{-0.66}$ & $-0.21^{+0.19}_{-0.48}$ & $0.08^{+0.34}_{-0.53}$ \\
\hspace{1mm}\\
$E_{22}$& $1.3 \pm 1.4$ & $2.1^{+1.8}_{-1.0}$ & $0.5^{+0.9}_{-1.5}$ & $0.9 ^{+1.2}_{-1.5}$ & $1.75 ^{+1.4}_{-1.0}$ & $0.4^{+0.8}_{-1.2}$ \\
\hspace{1mm}\\
$\mu_0 -1$& $0.05^{+0.22}_{-0.53}$ & $-0.09^{+0.15}_{-0.43}$ & $0.06^{+0.25}_{-0.45}$ & $0.05^{+0.23}_{-0.45}$ & $-0.15^{+0.13}_{-0.33}$ & $0.06^{+0.24}_{-0.37}$\\
\hspace{1mm}\\
$\eta_0 -1$& $0.96 \pm 1.1$ & $1.5^{+1.3}_{-0.8}$ & $0.3^{+0.6}_{-1.1}$ & $0.59 ^{+0.8}_{-1.0}$ & $1.22 ^{+0.96}_{-0.69}$ & $0.24^{+0.56}_{-0.83}$\\
\hspace{1mm}\\
$\Sigma_0 -1$& $0.36 \pm 0.18$ & $0.45^{+0.21}_{-0.17}$ & $0.12^{+0.09}_{-0.14}$ & $0.23^{+0.13}_{-0.15}$ & $0.28 ^{+0.13}_{-0.15}$ & $0.10^{+0.09}_{-0.10}$\\
\hspace{1mm}\\
\hline
\hspace{1mm}\\
$\Omega_{\rm b}h^2$& $0.02294 ^{+0.00049}_{-0.00063}$ & $0.02328 ^{+0.00048}_{-0.00063}$ & $0.02252 ^{+0.00036}_{-0.00043}$ & $0.02234 \pm 0.00025$ & $0.02244 \pm 0.00026$ & $0.02224 \pm 0.00024$ \\
\hspace{1mm}\\
$\Omega_{\rm c}h^2$ & $0.1202 \pm 0.0041$ & $0.1210 ^{+0.0041}_{-0.0046}$ & $0.1185 \pm 0.0039$ & $0.1184 \pm 0.0030$ & $0.1181 \pm 0.0030$ & $0.1173 \pm 0.0030$ \\
\hspace{1mm}\\
$H_0$ & $72.0 ^{+3.5}_{-4.8}$ & $74.7 ^{+3.5}_{-4.9}$ & $69.7 ^{+2.6}_{-3.2}$ & $67.6 \pm 1.6$ & $68.2 ^{+1.6}_{-1.8}$ & $67.1 \pm 1.5$\\
\hspace{1mm}\\
$\tau$ & $0.072 ^{+0.023}_{-0.026}$ & $0.073 \pm 0.024$ & $0.052 ^{+0.020}_{-0.025}$ & $0.059^{+0.018}_{-0.021}$ & $0.053 ^{+0.019}_{-0.021}$ & $0.044 ^{+0.016}_{-0.019}$\\ 
\hspace{1mm}\\
$n_s$ & $0.990 ^{+0.020}_{-0.025}$ & $1.004 ^{+0.019}_{-0.025}$ & $0.977 ^{+0.015}_{-0.017}$ & $0.9655 \pm 0.0097$ & $0.969 \pm 0.010$ & $0.9625 \pm 0.0092$\\
\hspace{1mm}\\
$\sigma_8$ & $0.827 ^{+0.033}_{-0.062}$ & $0.812 ^{+0.028}_{-0.054}$ & $0.808 ^{+0.035}_{-0.048}$ & $0.814 ^{+0.032}_{-0.049}$ & $0.788  ^{+0.022}_{-0.038}$ & $0.799 ^{+0.031}_{-0.037}$\\
\hspace{1mm}\\
$N_{\rm eff}$ & $3.41^{+0.36}_{-0.46}$ & $3.63^{+0.35}_{-0.48}$ & $3.19^{+0.30}_{-0.34}$ & $3.02 \pm 0.20$ & $3.06 \pm 0.21$ & $2.95 \pm 0.19$\\
\hspace{1mm}\\
\hline\hline
\end{tabular}}
\caption{Constraints at $68 \%$ c.l. on the cosmological parameters assuming modified gravity (parametrized by
$E_{11}$ and $E_{22}$) and varying the $6$ parameters
of the standard $\Lambda$CDM model plus $N_{\rm eff}$.}
\label{table5}
\end{center}
\end{table*}

\begin{table*}[t]
\begin{center}
\scalebox{1}{\begin{tabular}{ccccccc}
\hline\hline
\hspace{1mm}\\
& Planck TT & Planck TT + WL & Planck TT + lensing & Planck pol & Planck pol + WL & Planck pol + lensing \\
\hline\hline
\hspace{1mm}\\
$E_{11}$ & $0.08^{+0.36}_{-0.79}$ & $-0.21^{+0.20}_{-0.52}$ & $0.05^{+0.34}_{-0.57}$ & $0.06^{+0.34}_{-0.67}$ & $-0.25^{+0.20}_{-0.43}$ & $0.06^{+0.35}_{-0.53}$ \\
\hspace{1mm}\\
$E_{22}$& $1.2 ^{+1.4}_{-1.9}$ & $2.2^{+1.7}_{-1.1}$ & $0.5^{+0.9}_{-1.3}$ & $0.9 ^{+1.2}_{-1.6}$ & $1.8 ^{+1.3}_{-1.0}$ & $0.4^{+0.8}_{-1.2}$ \\
\hspace{1mm}\\
$\mu_0 -1$& $0.06^{+0.25}_{-0.55}$ & $-0.15^{+0.14}_{-0.37}$ & $0.03^{+0.24}_{-0.40}$ & $0.04^{+0.24}_{-0.46}$ & $-0.17^{+0.14}_{-0.30}$ & $0.04^{+0.24}_{-0.36}$\\
\hspace{1mm}\\
$\eta_0 -1$& $0.9 ^{+1.0}_{-1.3}$ & $1.6^{+1.2}_{-0.8}$ & $0.35^{+0.62}_{-0.94}$ & $0.6 ^{+0.8}_{-1.1}$ & $1.28 ^{+0.90}_{-0.69}$ & $0.28^{+0.58}_{-0.85}$\\
\hspace{1mm}\\
$\Sigma_0 -1$& $0.31 \pm 0.18$ & $0.38^{+0.20}_{-0.18}$ & $0.11^{+0.10}_{-0.13}$ & $0.22^{+0.13}_{-0.15}$ & $0.27 \pm 0.13$ & $0.10^{+0.09}_{-0.11}$\\
\hspace{1mm}\\
\hline
\hspace{1mm}\\
$\Omega_{\rm b}h^2$& $0.02267 ^{+0.00032}_{-0.00038}$ & $0.02281 ^{+0.00033}_{-0.00039}$ & $0.02238 \pm 0.00026$ & $0.02238 \pm 0.00018$ & $0.02243 \pm 0.00017$ & $0.02232 \pm 0.00017$ \\
\hspace{1mm}\\
$\Omega_{\rm c}h^2$ & $0.1170 \pm 0.0027$ & $0.1154 \pm 0.0024$ & $0.1171 \pm 0.0021$ & $0.1188 \pm 0.0016$ & $0.1180 \pm 0.0015$ & $0.1186 \pm 0.0015$ \\
\hspace{1mm}\\
$H_0$ & $68.8 ^{+1.3}_{-1.4}$ & $69.6 ^{+1.2}_{-1.3}$ & $68.5 \pm 1.0$ & $67.76 \pm 0.72$ & $68.12 \pm 0.70$ & $67.80 \pm 0.66$\\
\hspace{1mm}\\
$\tau$ & $0.068 ^{+0.022}_{-0.025}$ & $0.064 ^{+0.021}_{-0.025}$ & $0.051 ^{+0.019}_{-0.022}$ & $0.060^{+0.019}_{-0.022}$ & $0.054 ^{+0.020}_{-0.043}$ & $0.045 \pm 0.017$\\ 
\hspace{1mm}\\
$n_s$ & $0.9721 \pm 0.0076$ & $0.9765 \pm 0.0073$ & $0.9708 \pm 0.0064$ & $0.9665 \pm 0.0051$ & $0.9686 \pm 0.0051$ & $0.9669 \pm 0.0050$\\
\hspace{1mm}\\
$\sigma_8$ & $0.816 ^{+0.036}_{-0.059}$ & $0.784 ^{+0.022}_{-0.042}$ & $0.800 ^{+0.033}_{-0.038}$ & $0.815 ^{+0.033}_{-0.048}$ & $0.785  ^{+0.021}_{-0.034}$ & $0.803 ^{+0.030}_{-0.036}$\\
\hspace{1mm}\\
$\frac{dn_s}{dlnk}$ & $-0.0073^{+0.0097}_{-0.0086}$ & $-0.008^{+0.011}_{-0.009}$ & $0.0002 \pm 0.0079$ & $-0.0014 \pm 0.0072$ & $-0.0005 \pm 0.0070$ & $0.0016 \pm 0.0070$ \\
\hspace{1mm}\\
\hline\hline
\end{tabular}}
\caption{Constraints at $68 \%$ c.l. on the cosmological parameters assuming modified gravity (parametrized by
$E_{11}$ and $E_{22}$) and varying the $6$ parameters
of the standard $\Lambda$CDM model plus $dn_S/dlnk$.}
\label{table6}
\end{center}
\end{table*}

\begin{table*}[t]
\begin{center}
\scalebox{1}{\begin{tabular}{ccccccc}
\hline\hline
\hspace{1mm}\\
& Planck TT & Planck TT + WL & Planck TT + lensing & Planck pol & Planck pol + WL & Planck pol + lensing \\
\hline\hline
\hspace{1mm}\\
$E_{11}$ & $0.05^{+0.33}_{-0.71}$ & $-0.18^{+0.20}_{-0.52}$ & $0.05^{+0.35}_{-0.58}$ & $0.08^{+0.34}_{-0.68}$ & $-0.24^{+0.20}_{-0.44}$ & $0.05^{+0.34}_{-0.52}$ \\
\hspace{1mm}\\
$E_{22}$& $1.2 \pm 1.4$ & $2.1^{+1.6}_{-1.0}$ & $0.5^{+0.9}_{-1.4}$ & $0.6 ^{+0.8}_{-1.1}$ & $1.9 ^{+1.3}_{-1.0}$ & $0.4^{+0.8}_{-1.2}$ \\
\hspace{1mm}\\
$\mu_0 -1$& $0.04^{+0.24}_{-0.51}$ & $-0.13^{+0.14}_{-0.37}$ & $0.04^{+0.25}_{-0.41}$ & $0.06^{+0.24}_{-0.47}$ & $-0.17^{+0.14}_{-0.31}$ & $0.04^{+0.23}_{-0.36}$\\
\hspace{1mm}\\
$\eta_0 -1$& $0.9 ^{+1.0}_{-1.2}$ & $1.5^{+1.2}_{-0.8}$ & $0.36^{+0.62}_{-0.99}$ & $0.6 ^{+0.8}_{-1.1}$ & $1.30 ^{+0.91}_{-0.72}$ & $0.29^{+0.57}_{-0.83}$\\
\hspace{1mm}\\
$\Sigma_0 -1$& $0.31 \pm 0.16$ & $0.39^{+0.19}_{-0.15}$ & $0.11^{+0.09}_{-0.12}$ & $0.23^{+0.13}_{-0.16}$ & $0.29 \pm 0.13$ & $0.10^{+0.09}_{-0.11}$\\
\hspace{1mm}\\
\hline
\hspace{1mm}\\
$\Omega_{\rm b}h^2$& $0.02269 ^{+0.00041}_{-0.00046}$ & $0.02293 \pm 0.00042$ & $0.02248 \pm 0.00034$ & $0.02245 ^{+0.00024}_{-0.00026}$ & $0.02254 \pm 0.00023$ & $0.02236 \pm 0.00023$ \\
\hspace{1mm}\\
$\Omega_{\rm c}h^2$ & $0.1167 \pm 0.0028$ & $0.1147 \pm 0.0026$ & $0.1169 \pm 0.0023$ & $0.1187 \pm 0.0016$ & $0.1178 \pm 0.0015$ & $0.1185 \pm 0.0015$ \\
\hspace{1mm}\\
$H_0$ & $69.1 ^{+1.5}_{-1.7}$ & $70.2 ^{+1.5}_{-1.7}$ & $68.8 ^{+1.2}_{-1.4}$ & $67.98 ^{+0.84}_{-0.94}$ & $68.43 \pm 0.81$ & $67.92 \pm 0.77$\\
\hspace{1mm}\\
$\tau$ & $0.068 ^{+0.022}_{-0.024}$ & $0.066 ^{+0.021}_{-0.025}$ & $0.052 ^{+0.020}_{-0.023}$ & $0.061^{+0.020}_{-0.022}$ & $0.055 ^{+0.018}_{-0.022}$ & $0.046 \pm 0.018$\\ 
\hspace{1mm}\\
$n_s$ & $0.979 ^{+0.014}_{-0.016}$ & $0.988 \pm 0.015$ & $0.975 \pm 0.012$ & $0.9700 \pm 0.0086$ & $0.9732 \pm 0.0082$ & $0.9681 \pm 0.0080$\\
\hspace{1mm}\\
$\sigma_8$ & $0.816 ^{+0.033}_{-0.055}$ & $0.791 ^{+0.022}_{-0.043}$ & $0.803 ^{+0.035}_{-0.040}$ & $0.819 ^{+0.032}_{-0.050}$ & $0.788 ^{+0.021}_{-0.035}$ & $0.803 \pm 0.031$\\
\hspace{1mm}\\
$Y_P$& $0.258 \pm 0.023$ & $0.268 \pm 0.023$ & $0.253 \pm 0.021$ & $0.252 \pm 0.014$ & $0.254 \pm 0.013$ & $0.248 \pm 0.013$ \\
\hspace{1mm}\\
\hline\hline
\end{tabular}}
\caption{Constraints at $68 \%$ c.l. on the cosmological parameters assuming modified gravity (parametrized by
$E_{11}$ and $E_{22}$) and varying the $6$ parameters
of the standard $\Lambda$CDM model plus $Y_P$.}
\label{table7}
\end{center}
\end{table*}

\begin{figure}
\centering
\includegraphics[scale=0.7]{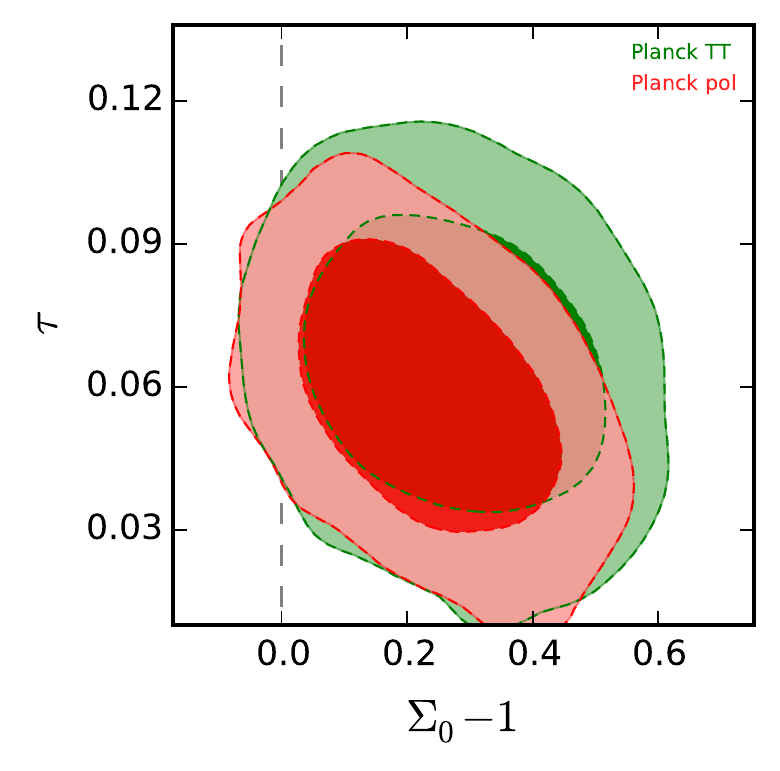}
\includegraphics[scale=0.7]{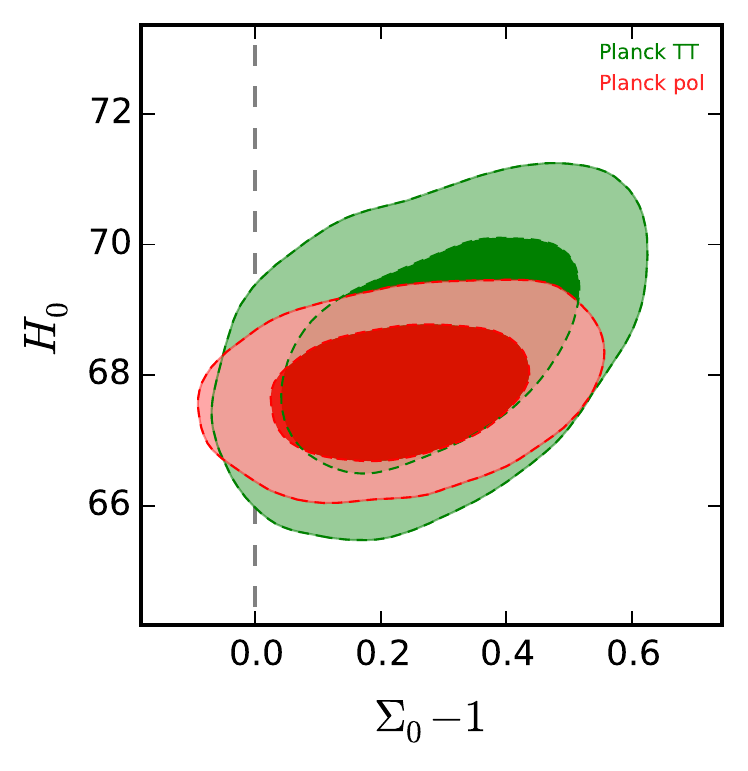}
\caption{Constraints at $68 \%$ and  $95 \%$ confidence levels on the $\Sigma_0-1$ vs $\tau$ plane (top panel) and on the
$\Sigma_0-1$ vs $H_0$ plane (bottom panel) from the Planck TT and Planck pol datasets. The $6$ parameters of the $\Lambda$CDM model are varied. Notice that $\Sigma_0$ is different from one (dashed vertical line) at about $95$ \% confidence level. A small degeneracy is present between $\Sigma_0$ and $\tau$: smaller optical depths are more compatible with the data if $\Sigma_0$ is larger than one (see top panel). Another degeneracy is present with the Hubble constant: larger values of the Hubble constant are more compatible with the considered data in case of $\Sigma_0$ different from one (bottom panel). }
\label{fig1}
\end{figure}

\begin{figure}
\centering
\includegraphics[scale=0.7]{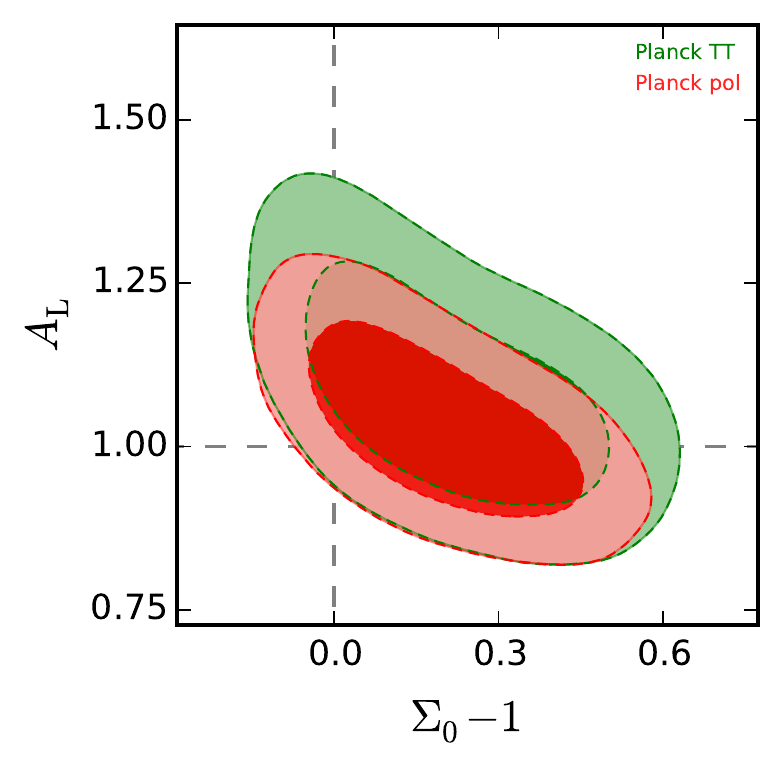}
\caption{Constraints at $68 \%$ and  $95 \%$ confidence levels on the $\Sigma_0-1$ vs $A_{\rm lens}$ plane from the 
Planck TT and Planck pol datasets. A strong degeneracy is present between $\Sigma_0$ and $A_{\rm lens}$: larger
values of $A_{\rm lens}$  are more compatible with the data if $\Sigma_0$ is smaller than one.}
\label{fig2}
\end{figure}

\begin{figure}
\centering
\includegraphics[scale=0.7]{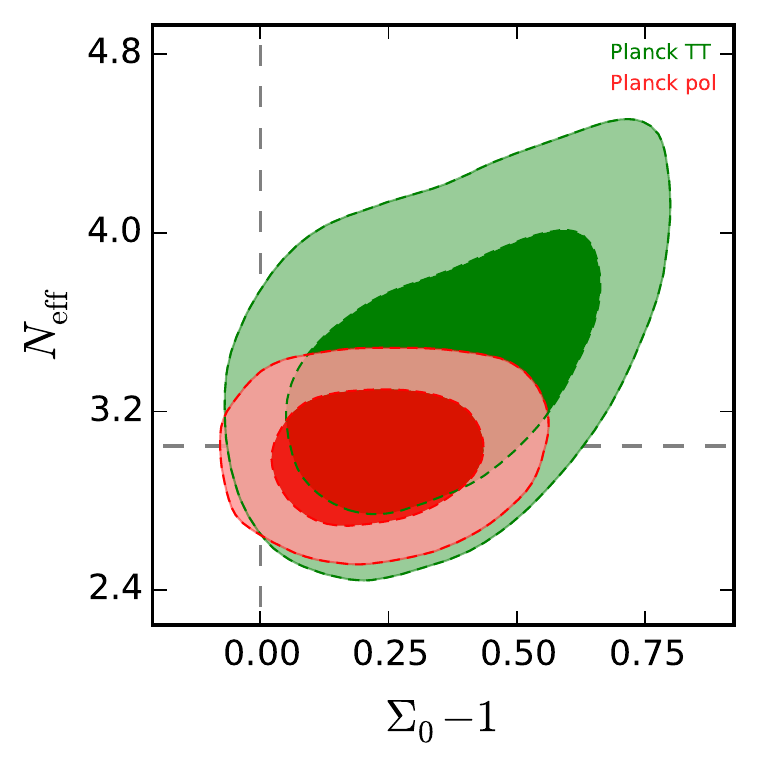}
\caption{Constraints at $68 \%$ and  $95 \%$ confidence levels on the $\Sigma_0-1$ vs $N_{\textrm{eff}}$ plane  from the Planck TT and Planck polarization datasets. Notice that $\Sigma_0$ is different from unity (dashed vertical line) at about the $95$ \% confidence level. 
A small direction of degeneracy is present between $\Sigma_0$ and $N_{\textrm{eff}}$: larger $N_{\textrm{eff}}$ are more compatible with the data if $\Sigma_0$ is larger than one in case of the Planck TT dataset.}
\label{fig3}
\end{figure}

\begin{figure}
\centering
\includegraphics[scale=0.7]{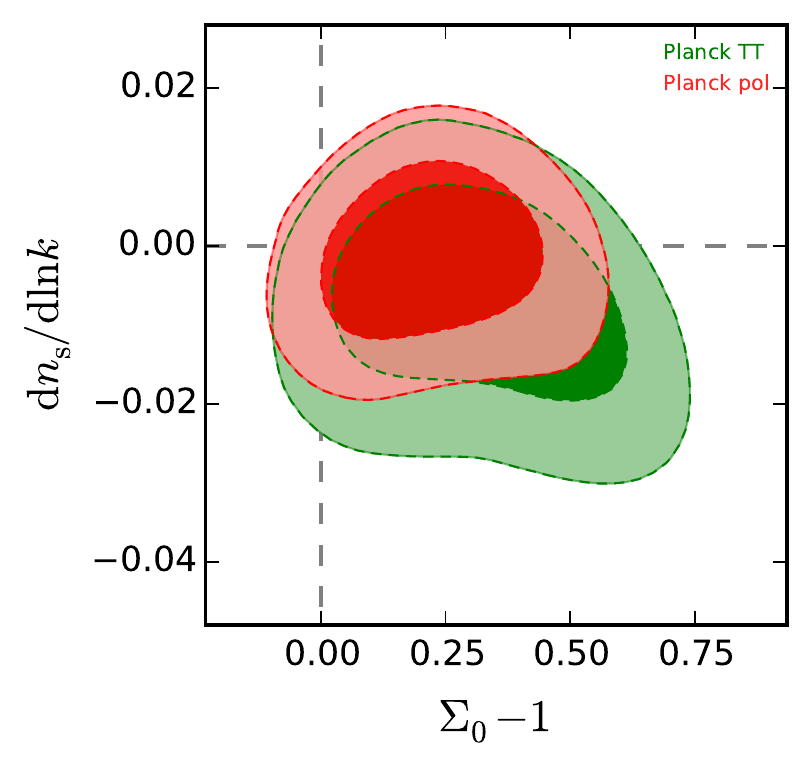}
\includegraphics[scale=0.7]{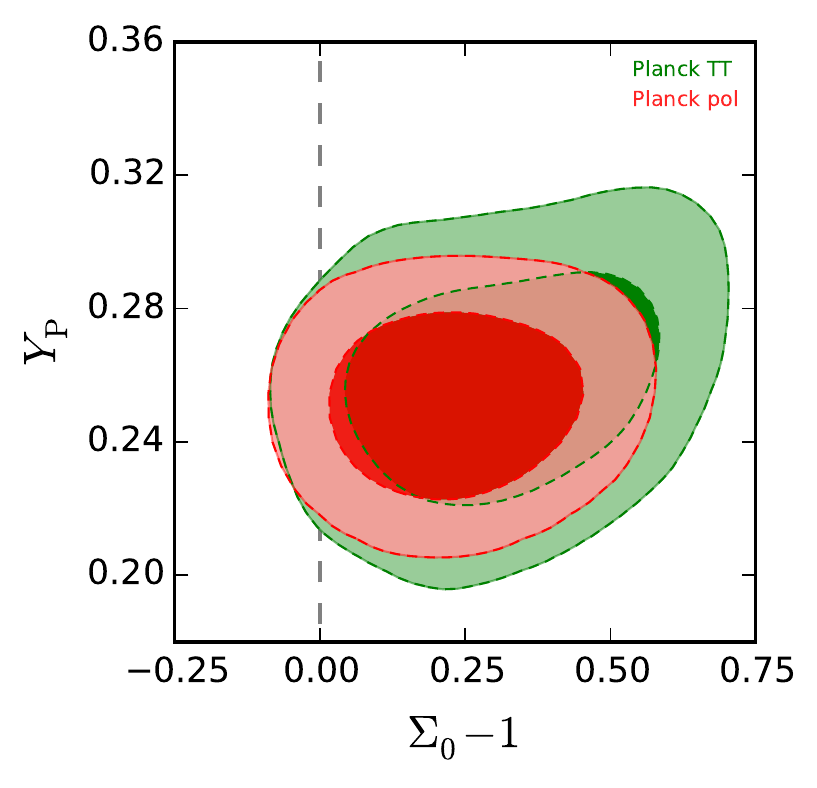}
\caption{Constraints at $68 \%$ and  $95 \%$ confidence levels on the $\Sigma_0-1$ vs $dn_s/dlnk$ plane (top panel) and on the $\Sigma_0-1$ vs $Y_p$ plane (bottom panel) from the Planck TT and Planck pol datasets. Notice that $\Sigma_0$ is different from unity (dashed vertical line) at about $95$ \% confidence level. There is virtually no degeneracy between $\Sigma_0$, the running of the scalar spectral index $dn_s/dlnk$ and the primordial helium abundance. }
\label{fig4}
\end{figure}

We constrain these cosmological parameters by using recent cosmological datasets.
First of all, we consider the full Planck 2015 release on temperature and polarization 
CMB angular power spectra, including the large angular scale temperature and polarization measurement by the Planck LFI
experiment and the small-scale temperature and polarization spectra by Planck HFI.
We refer to the Planck HFI small angular scale temperature data plus large angular scale
Planck LFI temperature and polarization data as {\it Planck TT}, while when we include small angular scale
polarization from Planck HFI as {\it Planck pol} (see \cite{plancklike2015}).
We also use  information on CMB lensing from Planck trispectrum data (see \cite{plancklens2015}) and we refer to this dataset
as {\it lensing}. Finally, we consider 
the weak lensing galaxy data from the CFHTlenS \cite{WL} survey with the priors and conservative cuts
to the data as described in \cite{planckparams2015} and we refer to this dataset as $WL$. 

To perform the analysis, we use our modified version, according to the Eqs.~\ref{DErel}, of the publicly available code \texttt{MGCAMB} \cite{Zhao:2008bn,Hojjati:2011ix} that modifies the original publicly code \texttt{CAMB} \cite{Lewis:1999bs} implementing the pair of functions $\mu(a,k)$ and $\eta(a,k)$,  as defined in \cite{Hojjati:2011ix}.
This code has been developed and tested in a completely independent way to the one used in \cite{Ade:2015rim}.

 We integrate \texttt{MGCAMB} in the latest July 2015 version of the publicly available Monte Carlo Markov Chain package \texttt{cosmomc} \cite{Lewis:2002ah} with a convergence diagnostic based on the Gelman and Rubin statistic. This version includes the support for the Planck data release 2015 Likelihood Code \cite{plancklike2015} (see \url{http://cosmologist.info/cosmomc/}) and implements an efficient sampling using the fast/slow parameter decorrelations \cite{Lewis:2013hha}.

\section{Results}

We first report the results assuming a modified gravity scenario parametrized by $\eta$ and $\mu$ and varying only 
the $6$ parameters of the standard $\Lambda$CDM model. The constraints on the several parameters are
reported in Table~\ref{table2}. When comparing the first and second column of our table, we see a complete agreement 
with the results presented in the first and third column of Table $6$ of \cite{Ade:2015rim}. Namely we find  evidence at $\sim 95 \%$ c.l. for $\Sigma_0-1$ different from zero for the Planck TT dataset, and  this indication is further confirmed when the WL dataset
is included. 

As fully discussed in \cite{plancklike2015}, the Planck polarization HFI data at small angular scales
fails to satisfy some of the internal checks in the data analysis pipeline. The results obtained by the inclusion 
of this dataset should therefore be considered as preliminary.
We report the constraints from the Planck pol dataset in columns $4$-$6$ in Table \ref{table2}.
As we can see, the small angular scale HFI polarization data improves the constraints on $\Sigma_0$,
also slightly shifting its value towards a better compatibility with standard $\Lambda$CDM. We
 can 
see however that the inclusion of small angular scale polarization does not alter substantially the conclusions
obtained when using just the Planck TT dataset.

Considering just the Planck TT dataset, it is interesting to note that in this modified gravity scenario,
the Hubble constant is constrained to be $H_0=68.5 \pm 1.1$ at $68 \%$ c.l., i.e. a value significantly larger than the
$H_0=67.3 \pm 0.96$ at $68 \%$ c.l. reported by the Planck collaboration assuming $\Lambda$CDM.
Combining the Planck TT dataset with the HST prior of $H_0=73.0\pm2.4$ from the revised analysis of 
\cite{riess} as in \cite{verde} we found
indeed an increased evidence for MG, with $\Sigma_0-1=0.33_{-0.15}^{+0.18}$ at $68 \%$ c.l..

Moreover, the amplitude of the r.m.s. mass density fluctuations $\sigma_8$ in our modified gravity scenario
is constrained to be $\sigma_8=0.816_{-0.052}^{+0.034}$ at $68 \%$ c.l., i.e. a value significantly weaker 
(and shifted towards smaller values) than the value of $\sigma_8=0.829\pm0.014$ at $68 \%$ c.l. reported by the 
Planck collaboration again under $\Lambda$CDM assumption. 

Considering the Planck pol dataset, the value of the optical depth is also significantly smaller in the MG scenario 
($\tau=0.059\pm0.020$ at $68 \%$ c.l.)  respect to the value obtained under standard $\Lambda$CDM model of 
$\tau=0.078\pm0.019$ at $68 \%$ c.l.,
i.e. reducing the tension with the Planck LFI large angular scale polarization constraint.
Interestingly, a smaller value for the optical depth of $\tau \sim 0.05$ is in better agreement with recent
optical and UV astrophysical data (see e.g. \cite{reio1,reio2,reio3}) and the reionization scenarios presented
in \cite{reiosimple}. A value of $\tau > 0.07$ could imply unexpected properties for high-redshift galaxies.
 Assuming an external gaussian prior of $\tau =0.05\pm0.01$ (at 68 $\%$ c.l..) as in \cite{reiosimple}
that would consider in a conservative way reionization scenarios where the star formation rate density rapidly declines
after redshift $z \sim 8$ as suggested by \cite{reio4}, 
we find that the Planck TT dataset provides the constraint $\Sigma_0-1=0.30\pm0.14$ at $68 \%$ c.l., i.e.
further improving current hints of  MG.
In this respect, future, improved, constraints on the value of $\tau$ from large-scale polarization measurements 
as expected from the Planck HFI experiment will obviously provide valuable information. 

The degeneracies between $\Sigma_0$, $H_0$ and $\tau$ can be clearly seen in Figure~\ref{fig1}
where we show the constraints at $68 \%$ and  $95 \%$ confidence levels on the $\Sigma_0-1$ vs $\tau$ plane (top panel) and on the $\Sigma_0-1$ vs $H_0$ plane (bottom panel) from the Planck TT and Planck pol datasets. As we can see, a degeneracy is present between $\Sigma_0-1$ and $\tau$: smaller optical depths are more compatible with the data if $\Sigma_0$ is larger than one (see top panel). As discussed, a second degeneracy is present with the Hubble constant: larger values of the Hubble constant are more compatible with the considered data in case of $\Sigma_0$ different from one (Bottom Panel).

As already noticed in \cite{Ade:2015rim} and as we will discuss in the next paragraph, the indication for MG
from the Planck data is strictly connected with the $A_{\rm lens}$ anomaly, i.e. with the fact that Planck
angular spectra show "more lensing" than expected in the standard scenario.
It is therefore not a surprise that when the Planck lensing data (obtained from a trispectrum analysis) that
is on the contrary fully compatible with the standard expectations is included in the analysis the
indication for modified gravity is significantly reduced to less than one standard deviation, as we can 
see from the third column of Table \ref{table2}.
On the other hand, when weak lensing data from the WL dataset is included, the indication for
MG increases, with $\Sigma_0-1$ larger than zero at more than $95 \%$ c.l..

In Tables~\ref{table3},~\ref{table5},~\ref{table6},~\ref{table7} we report constraints assuming one single parameter extension
to $\Lambda$CDM. In particular, we report constraints when adding as an extra parameter the
lensing amplitude $A_{\rm lens}$  (Table \ref{table3}),
the neutrino effective number $N_{\textrm{eff}}$  (Table \ref{table5}), the running of the scalar spectral index $dn_S/dlnk$ 
 (Table \ref{table6}) and, finally, the Helium abundance $Y_P$  (Table \ref{table7}).

As expected, there is a main degeneracy between the $A_{\rm lens}$ parameter and $\Sigma_0$, as
we can clearly see in Figure \ref{fig2} where we report the 2D posteriors at $68 \%$ and $95 \%$ c.l. 
in the $\Sigma_0-1$ vs $A_{\rm lens}$ plane from the Planck TT and Planck pol datasets. 
In practice, the main effect of a modified gravity model is to enhance the lensing signal in the
angular power spectrum. The same effect can be obtained by increasing $A_{\rm lens}$ and some
form of degeneracy is clearly expected between the two parameters. As we see from
the results in Table \ref{table3}, the value of the $A_{\rm lens}$ parameter, when MG is considered, is 
$A_{\rm lens}=1.09_{-0.13}^{+0.10}$, fully consistent with $1$, while for the standard $\Lambda$CDM
the constraint is $A_{\rm lens}=1.224_{-0.096}^{+0.11}$ at $68 \%$ c.l..
When also varying $A_{\rm lens}$ we found that the Planck pol datasets constraint the optical depth
to $\tau=0.056\pm0.020$ at $68 \%$ c.l.

On the other hand, by looking at the results in Tables~\ref{table5},~\ref{table6},~\ref{table7} we do not see a significant
degeneracy between the MG parameters and the new extra parameters.
A small degeneracy is however present between $\Sigma_0$ and the effective neutrino number $N_{\textrm{eff}}$.
We  see from Table \ref{table5}  that Planck TT data provides the constraint $N_{\textrm{eff}}=3.41_{-0.46}^{+0.36}$
at $68 \%$ c.l. that should be compared with  $N_{\textrm{eff}}=3.13_{-0.34}^{+0.30}$ at $68 \%$ c.l. from the 
same dataset but assuming the standard $\Lambda$CDM model. 
While the possibility of an unknown "dark radiation" component (i.e. $N_{\textrm{eff}}>3.046$, see e.g. \cite{DiValentino:2013qma,elena,DiValentino:2015wba}) is therefore more viable in 
a MG scenario, it is however important to note that when adding polarization data the constraint on the neutrino number is
perfectly compatible with the expectations of the standard three neutrino framework.
The constraints at $68 \%$ and $95 \%$ c.l. in the 
$\Sigma_0-1$ vs $N_{\textrm{eff}}$ planes are reported in Figure \ref{fig3}.

We also consider the possibility of a running of the scalar spectral index $dn_S/dlnk$. Results are reported in Table  \ref{table6}
and we find no degeneracy with MG parameters. The Planck TT constraint of $dn_S/dlnk=-0.0073_{-0.0086}^{+0.0097}$
at $68 \%$ c.l. is almost identical to the value $dn_S/dlnk=-0.0084\pm0.0082$ at $68 \%$ c.l. obtained using the same dataset but assuming standard $\Lambda$CDM.

We also considered variations in the primordial helium abundance $Y_P$ since it affects small angular scale anisotropies. Our results are in Table  \ref{table7}. The Planck TT constraint is found to be 
$Y_P=0.258\pm0.023$ at $68 \%$ c.l., slightly larger than the standard $\Lambda$CDM value of 
$Y_p=0.252\pm0.021$ at $68 \%$ c.l. obtained using the same dataset. While a larger helium abundance is
in better agreement with recent primordial helium measurements of \cite{izotov}, it is important to stress
that the inclusion of polarization yields a constraint that is almost identical to the one obtained under 
$\Lambda$CDM. The constraints at $68 \%$ and $95 \%$ c.l. in the $\Sigma_0-1$ vs $d n_S /dlnk$ and
$\Sigma_0-1$ vs $Y_P$ planes are reported in Figure \ref{fig4}.

\section{Conclusions}
\label{sec:conclusions}
  
In this paper, we have further investigated the current hints for a "modified gravity" scenario from the recent
Planck 2015 data release. We have confirmed that the statistical evidence for these hints, assuming the conservative
dataset of Planck TT, is, at most, at $\sim 95 \%$ c.l., i.e. not extremely significant. The statistical significance increases
when combining the Planck datasets with the WL cosmic shear dataset.
Indeed, the Planck dataset seems to provide lower values for the $\sigma_8$ parameter with respect to those
derived under the assumption of GR and $\Lambda$-CDM.

If  future astrophysical or cosmological measurements will point towards a lower value of the optical depth
of $\tau \sim 0.05$ or of the r.m.s. amplitude of mass fluctuations of $\sigma_8 \sim 0.78$ 
then the current hints for modified gravity could be further strenghtened.

However it also important to stress that when the CMB lensing likelihood is included in the analysis
the statistical significance for MG simply vanishes. 

We also investigated possible degeneracies with extra, non-standard parameters as the neutrino
effective number, the running of the scalar spectral index and the primordial helium abundance showing 
that the results on these parameters assuming $\Lambda$CDM are slightly changed when considering
the Planck TT dataset. Namely, under modified gravity we have  larger values for the neutrino
effective number, $N_{eff}=3.41_{-0.46}^{+0.36}$ at $68 \%$ c.l., and for the helium
abundance, $Y_p=0.258\pm0.023$. at $68 \%$ c.l..
However, the constraints on these parameters are practically identical those obtained under
GR when including the Planck HFI polarization data.

We have clearly shown that the slight Planck hints of  MG are strongly degenerate with the anomalous
lensing amplitude in the Planck CMB angular spectra parametrized by the $A_{\rm lens}$ parameter.
Indeed, the $A_{\rm lens}$ anomaly disappears when MG is considered. Clearly, undetected small
experimental systematics could be the origin of this anomaly. However our conclusions are that modified
gravity could provide a physical explanation, albeit exotic, for this anomaly that has been
pointed out already in pre-Planck CMB datasets \cite{preplanck}, was present in the Planck 2013 data release
\cite{planck2013} and seems still to be alive in the recent Planck 2015 release \cite{planckparams2015}
\footnote{Another possible physical explanation for the $A_{\rm lens}$ anomaly has been also very recently
proposed by \cite{julian} by considering the inclusion of compensated isocurvature perturbations.}.

An extra parameter we have not investigated here is the neutrino absolute mass scale $\Sigma m_{\nu}$.
Since MG is degenerate with the $A_{\rm lens}$ we expect that in a MG scenario current constraints
on the neutrino mass from CMB angular power spectra should be weaker. However a more detailed
computation is needed and we plan to investigate it in a future paper (\cite{future}).

During the submission process of our paper, another paper appeared \cite{pullen}, claiming
an indication for MG from cosmological data. The dataset used in that paper is completely independent
from the one used here and the MG parametrization is also different. Clearly a possible connection
between the two results deserves future investigation.

\section{Acknowledgments}
\label{sec:grazie}

It is a pleasure to thank Noemi Frusciante, Matteo Martinelli and Marco Raveri for useful discussions.
JS and EdV acknowledge support by ERC project 267117 (DARK) hosted by UPMC, and JS for support at JHU by National Science Foundation grant OIA-1124403 and by the Templeton Foundation. EdV has been supported in part by the Institute Lagrange de Paris. AM acknowledge support by the research grant Theoretical Astroparticle Physics number 2012CPPYP7 under the program
PRIN 2012 funded by MIUR and by TASP, iniziativa specifica INFN.

\end{document}